# Superconducting Electrometer Based on the Resistively Shunted Bloch Transistor

S. V. Lotkhov, H. Zangerle, A. B. Zorin, T. Weimann, H. Scherer, and J. Niemeyer

Physikalisch-Technische Bundesanstalt, Bundesallee 100, 38116 Braunschweig, Germany

*Abstract*—**We have fabricated the Bloch transistor shunted on-chip by a small-sized Cr resistor with $R_s \approx 1$ kΩ. The Bloch transistor normally consists of two small Josephson junctions connected in series, which in our case have been replaced by two superconducting interferometer loops, each with two junctions in parallel. A capacitively coupled gate is supplied to control the induced charge of the small intermediate electrode (island) of the transistor. The measured I-V curves show no hysteresis and correspond to the operation of a effective Josephson junction at the high-damping and strong-noise limits. The critical current of the system was found to be close to its nominal value, that is in accordance with the electromagnetic environment theory. The I-V curves were modulated by the gate with a period of e and a maximum swing of about 2 μV. Such rather moderate modulation results from the Josephson-to-charging energies ratio, $E_J/E_C \approx 9$, in our sample being far from its optimum value (≈ 0.3÷1).**

## I. INTRODUCTION

Superconducting circuits exhibiting the effects of tunneling of individual Cooper pairs in small tunnel junctions are being intensively studied both theoretically [1]–[4] and experimentally [3]–[10]. The most attractive feature of electronics based on single Cooper pair tunneling is the non-dissipative (Josephson) mechanism of charge transfer, which makes the dynamics of the circuits different from that of the normal single electron (SET) structures.

The peculiar behavior of superconducting circuits can be manifested in the Bloch transistor [2]. The Bloch transistor is the simplest superconducting device consisting of two ultrasmall tunnel junctions connected in series, with a gate capacitively coupled to the small intermediate electrode (island) of the transistor. In contrast to the traditional SET transistor [11], [12], the Bloch transistor is characterised by substantial Josephson coupling energy $E_J = (\Phi_0/2\pi)I_{AB}$. Here $\Phi_0 = h/2e$ is the flux quantum and $I_C^{AB} = \pi\Delta/2eR$ the Ambegaokar-Baratoff critical current at $T \ll T_c$, expressed via the superconducting energy gap $\Delta$ and the junction normal resistance $R$. Namely, $E_J$ in the Bloch transistor is comparable to the charging energy $E_c = e^2/2C$, where $C$ is the capacitance of the central island, and they both exceed the energy of thermal fluctuations $k_BT$. For Al junctions and typically $C \sim 0.5$ fF, the condition $E_J \sim E_c$ is fulfilled when the junction resistance $R$ is about several kΩ, i.e. of the same order of magnitude as the resistance quantum $R_Q = h/4e^2 \approx 6.45$ kΩ. It was predicted [2] and proved experimentally [6], [7] that the Bloch transistor can carry a finite supercurrent. The value of the critical current $I_c$ depends periodically on the polarization charge of the central island with a period of $2e$, i.e., it can be periodically modulated by the gate voltage $V_g$.

Due to $R \sim R_Q$, the Josephson phase is no longer a well-defined variable, and the critical current of these rather high-ohmic Josephson junctions is subject to fluctuations [13], [14] unless the environmental impedance and temperature are low. As a result for the unshunted Bloch transistor, the observable value of $I_c$ is considerably lower than the theoretical one [6], [7]. The fluctuations can, however, be reduced if one modifies the electromagnetic impedance $Z_e(\omega)$ seen by the device [3], [4], [15]. In particular, a resistive shunt $Z_e(\omega) = R_s \ll R_Q$ (i.e. purely ohmic impedance) can help to restore the critical current to be close to its theoretical value. This current can be effectively modulated by the gate. Moreover, if the self-capacitance of the shunt is sufficiently small, the I-V curve becomes non-hysteretic and its resistive branch can also be modulated by the gate: at fixed current $I_0 > I_c$, the voltage across the transistor appears as a periodical function of $V_g$. By analogy to well-known dc-SQUIDs, this property of the Bloch transistor can be advantageously used for electrometry, resulting in a high sensitivity of the device [15]. It is also important, that, because of $R_s \ll R_Q$, the bandwidth for operation frequencies available for the shunted Bloch transistor is substantially greater than that of an SET, whose tunnel junctions typically have a much higher resistance $R_T \gg R_Q$ [16].

In this paper we report on the first experimental results for the Bloch transistor with Al/AlO$_x$/Al small tunnel junctions, which is furnished with a Au/Cr on-chip shunt.

## II. THE EXPERIMENT

*A. Sample Design*

To make it possible to vary the $E_J/E_c$ ratio, we implemented the so-called SQUID-geometry: each transistor junction was replaced by two junctions in parallel (see the circuit diagramm and the SEM image in Figs. 1 and 2

Manuscript received September 15, 1998.

This work was supported in part by the EU (MEL ARI Research Project 22953 CHARGE).

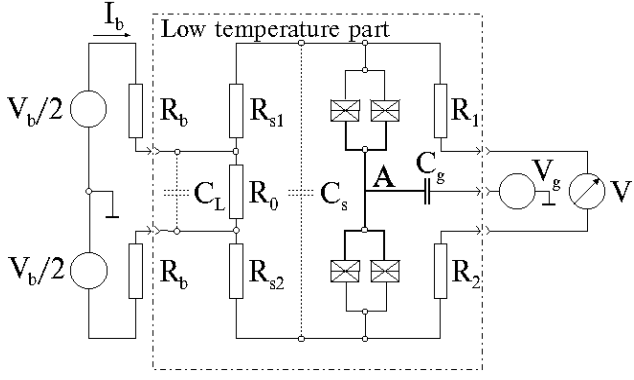

Fig. 1. The equivalent circuit of the shunted Bloch transistor. The crossed boxes denote Josephson junctions, the symbol "A" marks the central island. The parameters are $R_b = 100$ k$\Omega$, $R_1 = R_2 = R_{s1} + R_{s2} + R_0 = R_s = 1$ k$\Omega$, $C_g \approx 30$ aF, stray capacitances $C_L \sim 1$ nF and $C_s \sim 300$ aF.

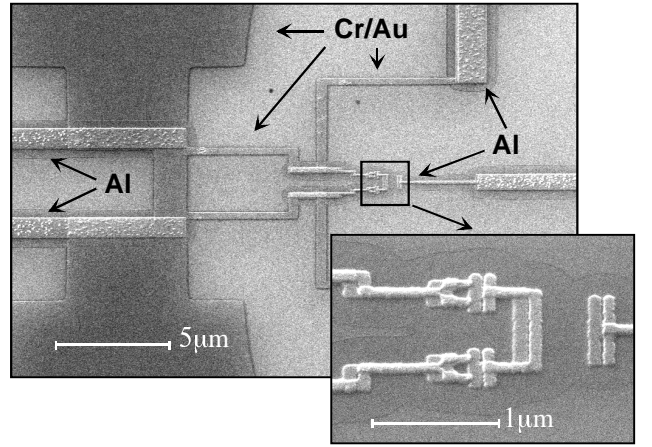

Fig. 2. The SEM image of the structure. The blow-up shows the area of the transistor itself. In each arm of the transistor, two junctions in parallel form an interferometer loop (so-called SQUID-geometry).

respectively), thus forming a small loop with the effective area of $A = 0.016$ µm$^2$ in between. An external perpendicular magnetic field, $B$, could tune the effective Josephson coupling $E_J = 2E_{J0}|\cos(\pi BA/\Phi_0)|$, where $E_{J0}$ is the Josephson coupling energy of a single junction on the assumption of their identity.

In order to achieve the purest resistive shunting of the transistor by a several-µm-long microstrip of resistance $R_s \approx 1$ k$\Omega$ and self-capacitance $C_s$ of up to several fF in the required frequency band $0 < \omega < \omega_J = 2eI_cR_s/\hbar$ (about several GHz) [15], [16] one has to reduce the contribution of the biasing leads of much larger capacitance $C_L$. That can be realized, for example, by using the special configuration for the shunting resistor (shown in Fig. 1) allowing injection of the bias current $I_b$ into a short section $R_0$ ($\approx 0.1R_s$) of the whole resistor $R_s$. That would necessarily require that 10 times larger biasing currents be supplied to the device, which in turn would lead to higher effective electron temperatures. In our structure, in order to reduce the electron temperature, the shunt was supplied by the large-area (about 0.07 mm$^2$) "cooling fins" placed at the points of current inlet [17]. The lines used for voltage measurement across the transistor were also equipped with on-chip resistors $R_{1,2}$ each of resistance $R_s$. Note that the whole resistor circuit occupied an area of 10 µm x 10 µm, which is much smaller than the characteristic wavelength $\lambda = 2\pi c/\omega_J \sim 0.1$ m.

*B. Sample Fabrication*

For fabrication of the sample (see the SEM-image in Fig. 2), we developed a two-level process with precise alignment of the resistive and superconducting layers. For both levels e-beam lithography followed by the "lift-off" procedure was used. First, the shunt with the cooling fins and additional resistors was formed. The mask for deposition of the metals (Cr, 15 nm / Au, 2 nm), used in this step, was the double layer PMMA/Copolymer, developed in the mixture of MIBK:IPA (1:2) followed by ECA:Ethanol(1:10). The resulting mask profile had an undercut which was useful for the cleaner lift-off process. The thin layer of Au over Cr was deposited in order to both ensure reliable electrical contact between Cr and Al layers and achieve the required specific film resistivities about 30 $\Omega$/square. In the second step, the Al-junctions were fabricated. Here we used the well-known double shadow angle-deposition technique with "mild" intermediate oxidation of the first Al-layer. In contrast to that, by fabrication of high-ohmic SET tunnel junctions (typically 100 k$\Omega$ per junction), the oxidation was carried out at a significantly lower pressure of O$_2$ (10$^{-3}$ mbar against several millibar for 20 min), the junction resistance being several k$\Omega$. The tri-level PMMA/Ge/Copolymer mask for Al-deposition was formed by dry etching with a larger undercut in the copolymer defining the hanging bridges of Ge at the places where the junctions would be formed as overlaps (see the blow-up at Fig. 2). The tunnel junctions were about 40 nm x 80 nm in size.

With the help of the transistor fabricated on the same chip with the same layout but without shunt we evaluated the electric parameters of the junctions and shunt as follows: $R \approx 4$ k$\Omega$ per junction, $\Delta \approx 175$ µeV, $E_J \approx 280$ µeV per SQUID, $C \sim 2.5$ fF, $E_c \sim 30$ µeV, $E_J/E_c \sim 9$, $R_s \approx 1$ k$\Omega$, and $R_0 \approx 0.1$ k$\Omega$. In order to estimate the capacitance $C$ we used the data obtained with the unshunted SET transistor having only one smaller junction in each arm and a total resistance of 18 kOhm. It was fabricated separately but experienced the same oxidation conditions for Al, and therefore nominally had the same thickness for the dielectric barrier. That is why we consider the charging energies to be proportional to the values of normal resistances for both samples and derived the aforementioned value of $E_c$ for the shunted transistor.

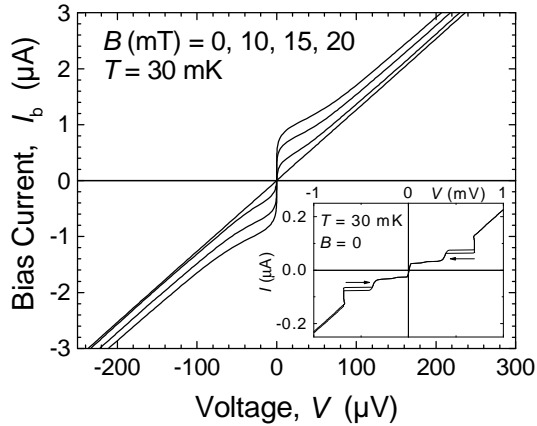
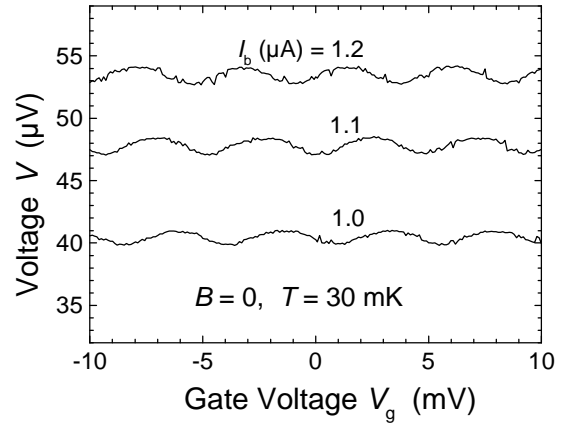

Fig. 3. The $I_b$-$V$ curves of the shunted Bloch transistor for several values of the magnetic field, corresponding to gradual suppression of the critical current. Note that the current flowing through the transistor is about $0.1 I_b$. The inset shows the $I$-$V$ curve of the similar, but unshunted, transistor fabricated on the same chip.

Fig. 4. The $V$-$V_g$ curves measured for several values of the bias current. Their period corresponds to the $e$-periodicity in the polarization charge of the island.

## III. RESULTS AND DISCUSSION

The measurements were performed in the dilution fridge at temperatures of 30 - 500 mK. The sample was placed into a shielding box and the microwave-frequency filtering of contact leads was done by Thermocoax® cable lines positioned in a low temperature section of the sample holder.

### A. I-V Curves

The inset in Fig. 3 shows the measured dc $I$-$V$ curve of the unshunted transistor and $I_b$-$V$ characteristics of the shunted devices. The former is characterized by the initial slope $dV/dI \approx 625\ \Omega$ of the so-called phase-diffusion branch [18] whose magnitude ($\pm 25$ nA) is substantially lesser than the AB critical current value $I_C^{AB} \approx 140$ nA [14] as well as the maximum critical current of the symmetric Bloch transistor [15], [19] with $E_J/E_c = 9$, which is accounted for by $I_C^{Bloch} \approx 0.7 \times I_C^{AB} \approx 100$ nA. On the contrary (Fig. 3), the $I_b$-$V$ curves of the shunted transistors exhibit the clear critical current behavior.

The characteristics of the shunted device show no hysteresis and can be effectively modulated by the magnetic field. Except the critical current "corners", their shape is close to the typical hyperbolic shape of the Josephson junction characteristics given by the Resistively Shunted Junction model at the limit of high damping (see, e.g. [20]). It is consistent with the upper estimation of the McCumber parameter for our structure ($C_s \approx 300$ aF, $R_s \approx 1$k$\Omega$): $\beta_C = (2\pi/\Phi_0) I_C^{Bloch} C_s R_s^2 \approx 0.1 < 1$. As we see from Fig. 3, the critical current "corners" of the curves are still considerably rounded. We associate this effect with the combined action of quantum and thermal fluctuations. Also we do not rule out some noise contrubution of the measuring setup. The maximum critical current value, $I_c \approx 90$-$100$ nA, can be extrapolated from the $I_b$-$V$ curves, taking into accountthe distribution of the bias current between the branch with the junctions and the branch with the resistors with the ratio: $R_0/(R_s - R_0) \approx 1:9$. One can see that, due to the resistive shunt, the measured magnitude of the critical current nearly reaches the nominal value $I_{CBloch}$. This effect stems from the classical dynamics of the overall Josephson phase in the high-damping limit, taking place in the case of resistive shunting of the transistor.

### B. Modulation Curves

At various fixed values of the bias current $I_b$ the gate modulation of voltage across the transistor was recorded (see Fig. 4). We observed the modulation period $1e$, that means that the parity effect in the superconducting island [21] did not manifest itself, as it was sometimes reported by other groups (see, e.g. [22]). (by us $1e$ periodicity has been observed for the unshunted transistor also.) Although of the fact, that individual quasiparticles were able to tunnel onto and out of the island, does not change the mechanism of supercurrent, it apparently reduces the amplitude of the gate modulation [6]. The maximum peak-to-peak modulation in our sample turned out to be some $2\ \mu V$, that is approximately half the value expected for $2e$-modulation for the given parameters of the transistor. As long as the shape of the modulation curves for high values of $E_J/E_c$ is close to the sin-shape, this reduction of the amplitude by one half agrees well with the theory.

Since the ratio $E_J/E_c \sim 9$ in our sample was far from its optimum value $\approx 0.3$-$1$ [15], [16], [19], we tried to reduce it by suppression of $E_J$ in the SQUIDs by magnetic field. (Note that the gate modulation of critical current for the optimised device can be as large as max($I_c$)/min($I_c$) ≈ 2.) Unfortunately, application of the field has not made the modulation distinctly deeper. We attribute this effect to the contributions of both suppression of the critical current and accompanying degradation of the sharpest slope (d$V$/d$I$) in the resistive branch due to the stronger effect of fluctuations.

The background charge noise measurement for our device was hindered by insufficient modulation and, hence, a small voltage-to-charge relation $|dV/dQ| \approx 5$ μV/$e$: the noise level of 25 nV/Hz$^{1/2}$ measured at $f = 10$ Hz corresponded to the noise of the test circuitry itself. This defines the charge sensitivity level for our device to be $5 \cdot 10^{-3}$ $e$/Hz$^{1/2}$.

To conclude, we designed, fabricated and measured a resistively shunted Bloch transistor. Owing to the on-chip shunt, we succeeded in observing the transistor's critical current which was close to its theoretical value. It was possible to periodically modulate the *I-V* characteristic of the transistor by the gate voltage. A relatively small modulation is explained by the fact that the Josephson-to-charging energies ratio was far apart from the optimum value. As a result, the actual charge sensitivity of the transistor as an electrometer was not high enough for measuring the background charge noise at the present level of noise of the test circuitry. Therefore, on the way to the practical Bloch electrometer the device parameters must be further optimised.

ACKNOWLEDGMENT

We thank V. A. Krupenin and A. B. Pavolotsky for useful discussions.